\documentclass[twoside,twocolumn,9pt]{article}
\usepackage{extsizes}
\usepackage[super,sort&compress,comma]{natbib}
\usepackage[version=3]{mhchem}
\usepackage[left=1.5cm, right=1.5cm, top=1.785cm, bottom=2.0cm]{geometry}
\usepackage{balance}
\usepackage{widetext}
\usepackage{times}
\usepackage{newtxtext,newtxmath}
\usepackage{eqnarray,amsmath}
\usepackage{sectsty}
\usepackage{graphicx}
\usepackage{booktabs}
\usepackage{lastpage}
\usepackage[format=plain,justification=raggedright,singlelinecheck=false,font={stretch=1.125,small,sf},labelfont=bf,labelsep=space]{caption}
\usepackage{float}
\usepackage{fancyhdr}
\usepackage{fnpos}
\usepackage[english]{babel}
\usepackage{array}
\usepackage{droidsans}
\usepackage{charter}
\usepackage[T1]{fontenc}
\usepackage[usenames,dvipsnames]{xcolor}
\usepackage{setspace}
\usepackage[compact]{titlesec}
\usepackage{bm}
\usepackage{multicol}

\usepackage{epstopdf}

\definecolor{cream}{RGB}{222,217,201}
\usepackage{graphicx}
\usepackage{xcolor}

\usepackage{amsfonts}
\usepackage{amsmath}
\usepackage{graphicx}
\usepackage{bm}
\def\red{\textcolor{red}} 

\def\Gr{{\rm \Gamma}}
\newcommand{\be}{\begin{equation}} 
\newcommand{\ee}{\end{equation}}
\newcommand{\bea}{\begin{eqnarray}}
\newcommand{\eea}{\end{eqnarray}}

\begin{document}

\makeFNbottom
\makeatletter

\def\red{\textcolor{red}}

\def\Gr{{\rm \Gamma}}

\renewcommand\LARGE{\@setfontsize\LARGE{15pt}{17}}
\renewcommand\Large{\@setfontsize\Large{12pt}{14}}
\renewcommand\large{\@setfontsize\large{10pt}{12}}
\renewcommand\footnotesize{\@setfontsize\footnotesize{7pt}{10}}
\makeatother

\renewcommand{\thefootnote}{\fnsymbol{footnote}}
\renewcommand\footnoterule{\vspace*{1pt}%
\color{cream}\hrule width 3.5in height 0.4pt \color{black}\vspace*{5pt}}
\setcounter{secnumdepth}{5}

\makeatletter
\renewcommand\@biblabel[1]{#1}
\renewcommand\@makefntext[1]%
{\noindent\makebox[0pt][r]{\@thefnmark\,}#1}
\makeatother
\renewcommand{\figurename}{\small{Fig.}~}
\sectionfont{\sffamily\Large}
\subsectionfont{\normalsize}
\subsubsectionfont{\bf}
\setstretch{1.125} 
\setlength{\skip\footins}{0.8cm}
\setlength{\footnotesep}{0.25cm}
\setlength{\jot}{10pt}
\titlespacing*{\section}{0pt}{4pt}{4pt}
\titlespacing*{\subsection}{0pt}{15pt}{1pt}

\fancyfoot{}
\fancyfoot[LE]{\footnotesize{\sffamily{\hspace{2pt}\thepage}}}
\fancyfoot[RO]{\footnotesize{\sffamily{\hspace{2pt}\thepage}}}

\fancyhead{}
\renewcommand{\headrulewidth}{0pt}
\renewcommand{\footrulewidth}{0pt}
\setlength{\arrayrulewidth}{1pt}
\setlength{\columnsep}{6.5mm}
\setlength\bibsep{1pt}

\makeatletter
\newlength{\figrulesep}
\setlength{\figrulesep}{0.5\textfloatsep}

\newcommand{\topfigrule}{\vspace*{-1pt}%
\noindent{\color{cream}\rule[-\figrulesep]{\columnwidth}{1.5pt}} }

\newcommand{\botfigrule}{\vspace*{-2pt}%
\noindent{\color{cream}\rule[\figrulesep]{\columnwidth}{1.5pt}} }

\newcommand{\dblfigrule}{\vspace*{-1pt}%
\noindent{\color{cream}\rule[-\figrulesep]{\textwidth}{1.5pt}} }

\makeatother

\twocolumn[
  \begin{@twocolumnfalse}
\sffamily
\begin{tabular}{m{0.0cm} p{16.0cm} }

\quad &
\noindent\LARGE{\textbf{Empty liquid state and re-entrant phase behavior of the patchy colloids confined in 
porous media}} \\
\vspace{0.3cm} & \vspace{0.3cm} \\
 & \noindent\large{Taras Hvozd$^1$, Yurij V. Kalyuzhnyi$^1$, Vojko Vlachy$^2$, Peter T. Cummings$^3$} \\
 \quad & \textit{$^1$~Institute for Condensed Matter Physics of the National Academy of Sciences of Ukraine 1 Svientsitskii St., Lviv, Ukraine 79011;} \\ 
 &\textit{ E-mail: tarashvozd@icmp.lviv.ua,$\;\;$yukal@icmp.lviv.ua} \\ 
  \quad & \textit{$^2$~Faculty of Chemistry and Chemical Technology, University of Ljubljana, Ve\v{c}na pot 113, SI--1000 Ljubljana, Slovenia. } \\ 
 &\textit{E-mail: vojko.vlachy@fkkt.uni-lj.si} \\ 
\quad & \textit{$^3$~Department of Chemical and Biochemical Engineering,
  Vanderbilt University, Nashville, TN 37235-1604, USA;} \\ 
 &\textit{E-mail: peter.cummings@Vanderbilt.Edu } \\ 
 \vspace{0.3cm} & \vspace{0.3cm} \\
\quad 
& \noindent\normalsize
{Patchy colloidal model with three and four equivalent patches, confined in the attractive random porous media, undergo re-entrant gas-liquid phase separation with the possibility for the liquid phase density to approach zero. This unusual behavior is caused by an interplay between strong fluid-fluid bonding interactions and weak fluid-matrix attractions.
At high temperature the shape of the phase diagram is determined by the attractive interaction between the fluid particles;  weak Yukawa attraction between fluid and obstacles only slightly enhances the fluid-fluid bonding. At low enough temperature the network of the fluid particles is formed  and the shape of the phase diagram becomes defined by the Yukawa fluid-obstacle attraction. Due to this interaction a layer of mutually bonded particles around the obstacles is formed and the 
network becomes fluid particles in the network is defined by a spatial arrangement of the matrix particles. 
These features may open a new possibilities in making the equilibrium gels with the predefined nonuniform distribution of the particles.}
\end{tabular}

 \end{@twocolumnfalse} \vspace{0.6cm}

]

\renewcommand*\rmdefault{bch}\normalfont\upshape
\rmfamily
\section*{}
\vspace{-1cm}

The concept of the ``empty liquid state'' has been formulated over a decade ago \cite{Bianchi2006} and since then a substantial amount of work has been published to identify and rationalize this phenomenon in framework of the patchy colloidal models  
\cite{Tavares2009,Russo2011a,Russo2011b,Heras2011,Kalyuzhnyi2013,Rovigatti2013,Sokolowski2014,Tavares2017}. 
According to Bianchi et al., \cite{Bianchi2006} a decrease of the valency of the patchy colloids (number of singly bondable patches) causes the liquid-gas phase diagram to shrink and disappear in the limit of two patches per particle. Upon approaching this limit the critical  temperature decreases, the liquid branch of the phase diagram is moving towards the gas branch and  the liquid phase density can take arbitrarily small values. This state of the liquid phase was named ``empty'' \cite{Bianchi2006}. To maintain continuous change of the valency, the authors consider a two-component mixture of patchy hard spheres with two and three bonding sites and assume that the phase behavior of this mixture can be described by a pseudo one-component fluid with an effective valency, which is defined by the relative concentration of the components. Subsequently, it 
was realized that the empty liquid state can be achieved assuming bonding sites to be of the different
type. Due to the competition between formation of the bonds connecting different patches the particles can self-assembly into the structures of different type, e.g. chains and branched structures. Appropriate choice for the energy parameters, which describe interaction between different patches, makes chain formation preferable at low  and branching at higher temperatures; as a result the empty liquid state can be achieved 
\cite{Russo2011a,Russo2011b,Heras2011,Kalyuzhnyi2013,Rovigatti2013,Sokolowski2014,Tavares2017}. 
The corresponding phase diagrams are re-entrant, which is rather unusual feature for the one-component fluids and makes them different in comparison with those observed earlier \cite{Bianchi2006}. 
The ability of liquid to be stabilized at vanishing low density offers the possibility
to establish equilibrium colloidal gel state \cite{Sciortino2017}. In this state the system is thermodynamically stable, in contrast to the standard colloidal gels, the equilibrium gel is reversible and does not age. Recently, the empty liquid state was observed experimentally for several systems, including suspensions of laponit \cite{Ruzicka2011a,Ruzicka2011b} and montmorillonite \cite{Pujala2015,Pajula2019} clay particles (see \cite{Sciortino2017} and references therein).

In this Letter we have studied the phase behavior of the patchy colloidal particles with several equivalent singly bondable patches, confined in the attractive porous media. According to our study, confinement give rise to the re-entrant phase diagram with arbitrarily low density of the liquid phase. 

\begin{figure}[!h]
	\centering
	\includegraphics[clip,height=0.25\textwidth,angle=0]{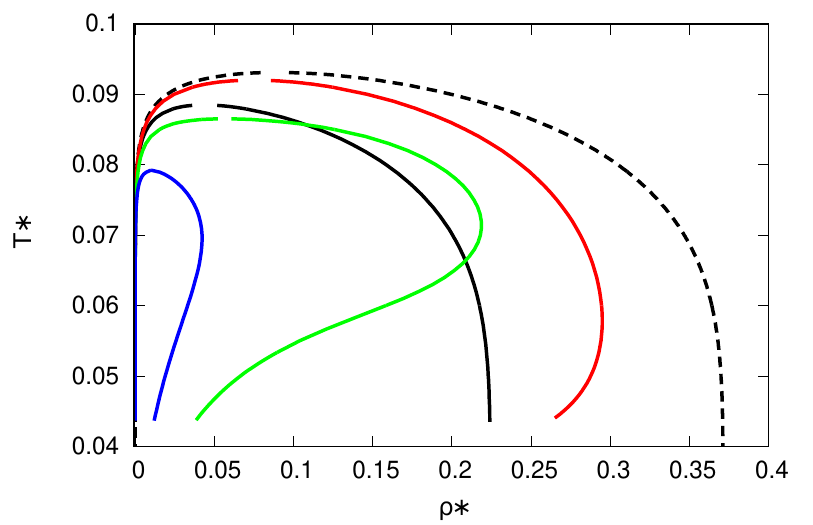}\\
	\includegraphics[clip,height=0.25\textwidth,angle=0]{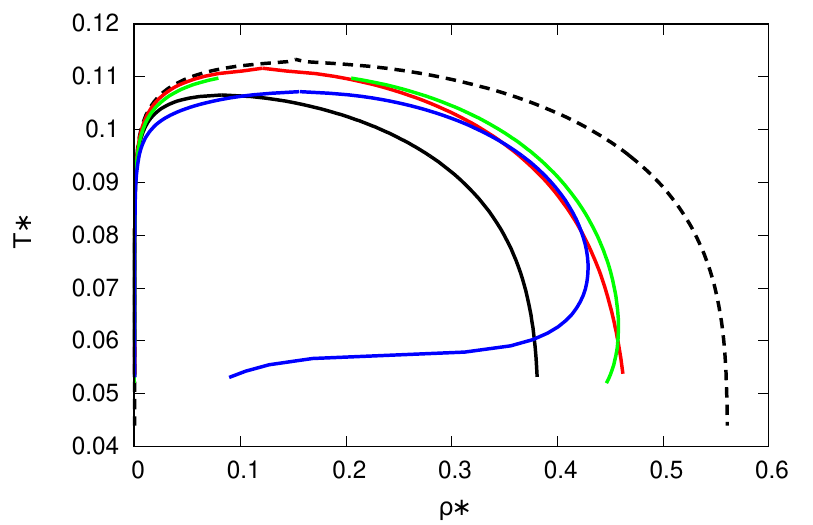}
	\caption{\label{fig1} Liquid-gas phase diagrams in $T^*$ vs $\rho_1^*$ coordinate frame for the three-patch 
(top panel) and 
four-patch (low panel) colloidal model confined in the Yukawa hard-sphere matrix with $\eta_0=0.1$ and for $\epsilon_{01}^{(Y)}/\epsilon_{11}^{(as)}=0.0$
(black lines), $\epsilon_{01}^{(Y)}/\epsilon_{11}^{(as)}=0.1$ (red lines), 
$\epsilon_{01}^{(Y)}/\epsilon_{11}^{(as)}=0.16$ (green lines),
$\epsilon_{01}^{(Y)}/\epsilon_{11}^{(as)}=0.22$
(blue lines). Here black dashed lines denote the phase diagram at $\eta_0=0$ and $\epsilon_{01}^{(Y)}/\epsilon_{11}^{(as)}=0.0$. }
\end{figure}

{\it Model.} -- We consider one-component patchy hard-sphere fluid confined in the random porous media. The media is represented by the matrix of Yukawa hard spheres quenched at hard-sphere fluid equilibrium. Each particle of the fluid has $n_s$ square-well bonding sites (patches) located on its surface. Interaction between the fluid particles and between matrix and fluid particles is described by the following pair potentials:
\be
U_{11}(12)=U_{11}^{(hs)}(r)+\sum_{KL}U_{KL}^{(as)}(12)
\label{U11}
\ee
and
\be
U_{01}(r)=U_{10}(r)=U_{01}^{(hs)}(r)-{\epsilon_{01}^{(Y)}\sigma_{01}\over r}
\exp{\left[-\alpha\left(r-\sigma_{01}\right)\right]},
\label{U01}
\ee
respectively. Here
\be
U^{(as)}_{KL}(12)=U^{(as)}_{11}(z_{12})=\left\{
\begin{array}{rl}
-\epsilon^{(as)}_{11}, & {\rm for}\;z_{12}\le\omega\\
0, & {\rm otherwise}
\end{array}
\right.,
\label{UKL}
\ee
the lower indices 0 and 1 denote the particles of the matrix and fluid, respectively.
1 and 2 denote position and orientation of the corresponding particles, the lower indices
$K$ and $L$ take the values $A,B,C,\ldots$ and denote bonding sites. $U^{(hs)}_{11}(r)$ and 
$U^{(hs)}_{01}(r)$ are fluid-fluid and fluid-matrix hard-sphere potentials, respectively,
$\epsilon^{(Y)}_{01}$ is the strength of Yukawa interaction, $z_{12}$ denotes the distance
between corresponding bonding sites, while $\epsilon_{11}^{(as)}$ and $\omega$ are the depth and
width of the site-site square-well interaction. The hard-sphere sizes of the fluid and matrix 
particles are $\sigma_1$ and $\sigma_0$, respectively, $\sigma_{01}=(\sigma_1+\sigma_0)/2$
and we consider the system at temperature $T$, having the fluid and matrix number densities $\rho_1$ and $\rho_0$.

{\it Theory.} -- Theoretical description of the model is carried out combining Wertheim's
thermodynamic perturbation theory \cite{Wertheim1987} (TPT), scale particle theory 
\cite{Holovko2009,Patsahan2011,Holovko2013}  (SPT) and 
replica Ornstein-Zernike (ROZ) equation \cite{Given1992,Given1993}. According to Wertheim's TPT 
\cite{Wertheim1987} Helmholtz free energy of the system is
\be
A=A_{hs}+\Delta A_{Y}+\Delta A_{as},
\label{TPT}
\ee
where $A_{hs}$ is Helmholtz free energy of the hard-sphere fluid confined in the hard-sphere
matrix, while $\Delta A_{Y}$ and $\Delta A_{as}$ are contributions to the free energy due to Yukawa
and bonding interactions, respectively. $A_{hs}$ is calculated using extension of the SPT, 
which provides analytical expressions for Helmholtz free energy, chemical potential 
and pressure of the hard-sphere fluid adsorbed in the hard-sphere matrix 
\cite{Holovko2009,Patsahan2011,Holovko2013,Kalyuzhnyi2014} (see Appendix A for details). 
$\Delta A_{Y}$ is calculated using numerical solution of the ROZ equation \cite{Given1992,Given1993} for the hard-sphere fluid in the Yukawa hard-sphere matrix. Corresponding
thermodynamic properties are calculated using the energy route. Solution
of the ROZ equation is obtained combining Percus-Yevick (PYA) and Mean Spherical (MSA) 
approximations (see Appendix B for details). For 
$\Delta A_{as}$ we have \cite{Wertheim1987}
\be
{\beta\Delta A_{as}\over V}=n_s\rho_1\left(\ln{X}-{1\over 2}X+{1\over 2}\right),
\label{Aas}
\ee
where $X=\left(\sqrt{1+2\Delta}-1\right)/\Delta$, with 
$
\Delta=8\pi n_s\rho_1g_{11}(\sigma_{1})\int_{\sigma_1}^{\sigma_1+\omega}{\bar f}(r)r^2dr, 
$ 
and
\be
{\bar f}(r)=\left(e^{\beta\epsilon}-1\right)\left(\omega+\sigma_1-r\right)^2
\left(2\omega-\sigma_1+r\right)/\left(6\sigma_1^2r\right)
\label{fbar}
\ee
and $g_{11}(r)$ is the radial distribution function (RDF) of the 
hard-sphere fluid confined in the Yukawa hard-sphere matrix. This RDF follows from the solution of 
the ROZ equation.

{\it Results.} -- The phase behavior of the model in question was investigated for the three-
and four-patch colloids ($n_s=3,4$) being trapped in the Yukawa hard-sphere matrix. The fluid and matrix particles had equal diameters
, i.e. $\sigma_0=\sigma_1=\sigma$. Matrix packing fraction was $\eta_0=\pi\rho_0\sigma_0^3/6=0.1$, width of the square-well site-site potential $\omega=0.119\sigma$, and we examined three different values for the strength of the Yukawa interaction 
$\epsilon_{01}^{(Y)}/\epsilon_{11}^{(as)}=0.0, 0.1, 0.16$, and $0.22$.

\begin{figure}[!h]
	\centering
	\includegraphics[clip,height=0.25\textwidth,angle=0]{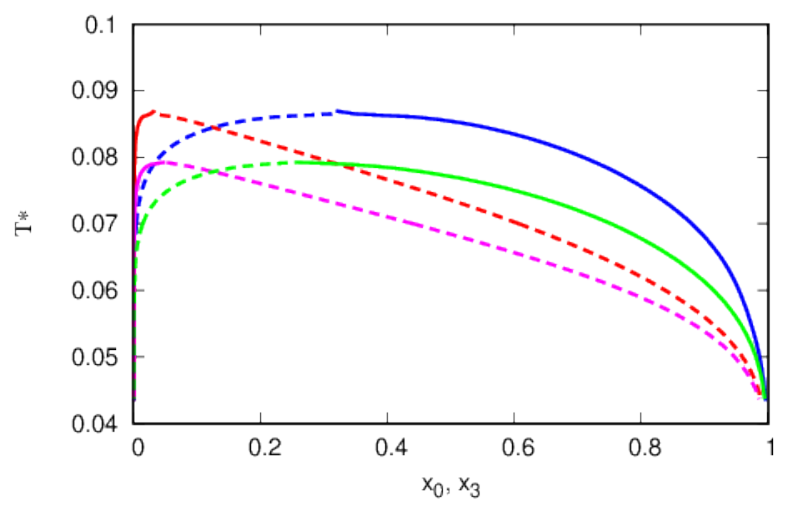}
	\caption{\label{fig2} Fractions of free $x_0$ (red and pink lines) and $3$-times bonded 
(blue and green lines) particles along
the binodals for the model with three patches and with $\epsilon_{01}^{(Y)}/\epsilon_{11}^{(as)}=0.16$
(lower set of the curves) and
$\epsilon_{01}^{(Y)}/\epsilon_{11}^{(as)}=0.22$ (upper set of the curves). Here dashed lines denote the gas phase and solid lines 
denote the liquid phase. }
\end{figure}

Our results for the phase behavior of the model system with $n_s=3$ and $n_s=4$ are shown in Fig. \ref{fig1}. For the reference we also included results for the phase diagrams 
in case that no obstacles are present ($\eta_0=0$, black dashed curves). The phase diagrams 
of the model liquids, confined in purely repulsive hard-sphere matrix ($\epsilon_{01}^{(Y)}$=0, black solid curves), 
are almost twice narrower than the $\eta_0=0$ cases, while their critical points move toward lower densities and 
temperatures.
For sufficiently low temperature the density of the liquid phase becomes temperature-independent, reaching a certain limiting value. At this temperature almost all bonding sites of the fluid particles are occupied and further decrease of the temperature does not change much the internal structure of the network formed. Such features of the phase diagram 
have already been observed before \cite{Bianchi2006,Kalyuzhnyi2014}. 

Weak Yukawa attraction  ($\epsilon_{01}^{(Y)}/\epsilon^{(as)}_{01}=0.1$) enhances the bonding and the phase diagrams become wider then at $\epsilon_{01}^{(Y)}=0.0$. In the same time, the critical points move towards higher 
temperatures and densities (note the red curves on this Figure). Such behavior is consistent with that observed in case of a simple fluid, confined in the attractive matrix \cite{Nelson2020}. However, the overall shape of the phase diagram shown in Fig. 
1 is different then observed before; at low temperature the liquid branch is moving towards the lower densities, clearly demonstrating re-entrant behavior. 

Moreover,  for even stronger Yukawa attraction ($\epsilon_{01}^{(Y)}/\epsilon_{11}^{(as)}=0.16$ and 0.22; green and
blue curves, respectively), the phase diagrams show re-entrant behavior at much higher temperatures. 
At the same time the width of the phase coexistence regions reduces substantially, especially at higher 
values of $\epsilon_{01}^{(Y)}$ ($\epsilon_{01}^{(Y)}/\epsilon_{11}^{(as)}=0.22$, blue curves). While for  
$\epsilon_{01}^{(Y)}/\epsilon_{11}^{(as)}=0.16$ and $n_s=3$ (green curves) the phase diagram in the 
temperature range ~ $0.065\leq T^* \leq 0.087$ is wider then the corresponding phase diagram for the
model with hard-sphere matrix ($\epsilon_{01}^{(Y)}=0$), the phase diagram for 
$\epsilon_{01}^{(Y)}/\epsilon_{11}^{(as)}=0.22$ and $n_s=3$ (blue curve) is much narrower and located 
completely inside
the phase diagram for the model with $\epsilon_{01}^{(Y)}=0$. Similar behavior is observed also for the four-patch version of the model. 
Thus, at sufficiently high values of the Yukawa attraction phase diagrams for the models examined here 
exhibit the re-entrant behaviour, eventually reaching the empty liquid state. The shape of the phase diagrams, shown in Fig. 1, is 
similar as observed at bulk conditions for the models with patches of different type
\cite{Russo2011a,Russo2011b,Kalyuzhnyi2013,Rovigatti2013,Sokolowski2014}. However, the underlying physics of this
behavior is different. In the studies mentioned above, the re-entrant phase diagram is a consequence of the
competition between chain formation and branching. 
At higher temperatures (but still below the critical point) the particles are connected by the 
3d network of bonds, which results in the phase separation. Upon the temperature decrease, the network
is destroyed in favour of chains and at zero temperature the phase separation is suppressed. 

\begin{figure}[!h]
	\centering
	\includegraphics[clip,height=0.3\textwidth,angle=0]{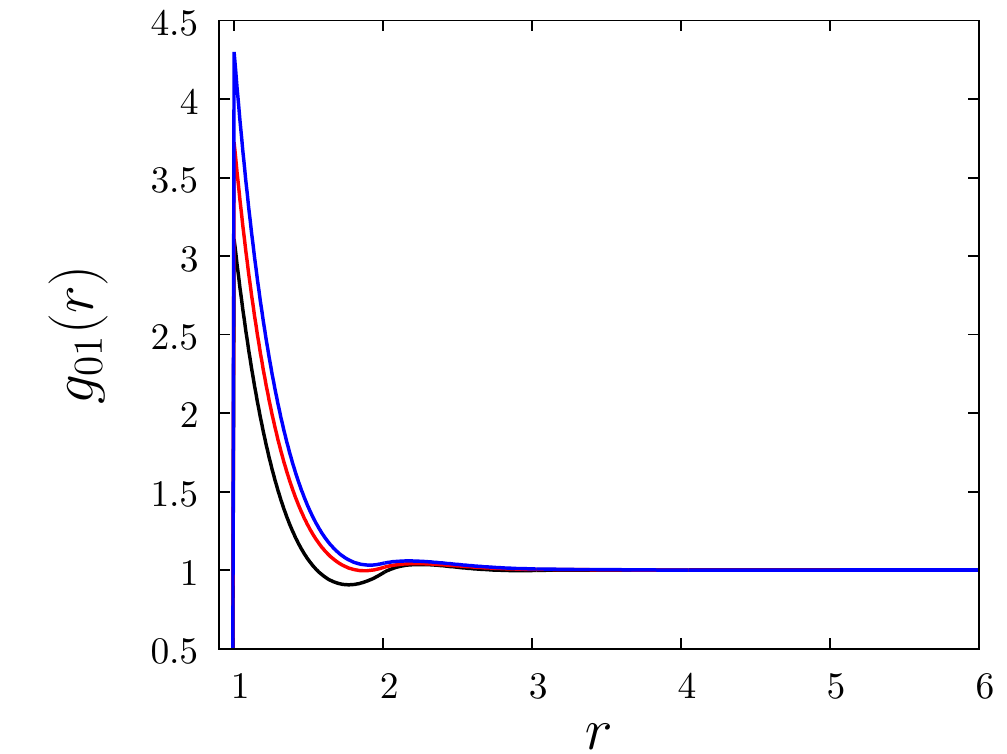}\\
	\includegraphics[clip,height=0.3\textwidth,angle=0]{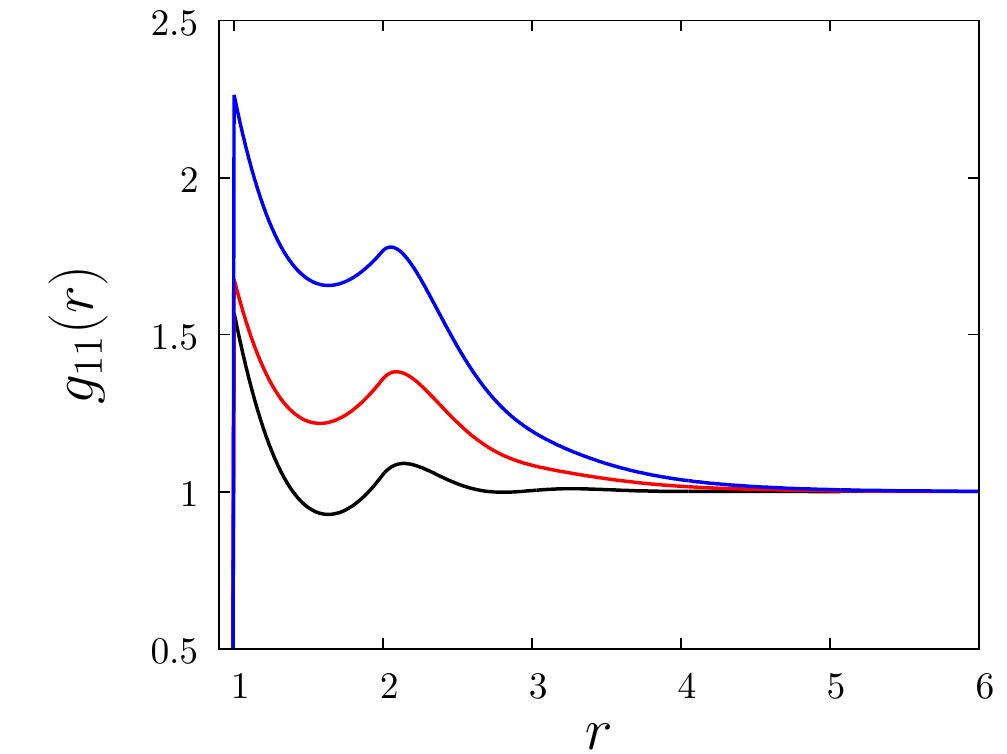}
	\caption{\label{fig3} Radial distribution functions $g_{01}(r)$ (top panel) and $g_{11}(r)$ (low panel)
for the hard-sphere fluid confined in Yukawa hard-sphere matrix with 
$\epsilon_{01}^{(Y)}/\epsilon_{11}^{(as)}=0.16$ and $\eta_0=0.1$ along the liquid branch of the phase diagram at 
$\rho_1^*=0.22,\;T^*=0.07$ (black lines), $\rho_1^*=0.06,\;T^*=0.049$ (red lines) and 
$\rho_1^*=0.027,\;T^*=0.04$ (blue lines). }
\end{figure}

\begin{figure}[!h]
	\centering
	\includegraphics[clip,height=0.25\textwidth,angle=0]{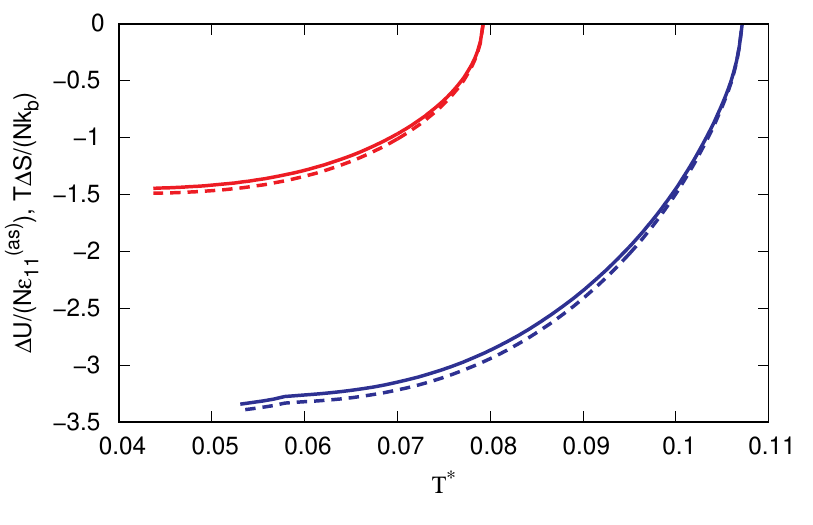}
	\caption{\label{fig4} Differences in the internal energies $\Delta U/(N\epsilon_{11}^{(as)})$ (solid lines) and entropies 
$T^*\Delta S/(Nk_B)$ (dashed lines) of the liquid and gas phases along the binodals for the models with three (red lines) and four (blue lines) patches at $\epsilon^{(Y)}_{01}/\epsilon^{(as)}_{11}=0.22$. }
\end{figure}

In contrast to this, the 3d network of the model particles  does not experience the restructuring of the  type mentioned above. By lowering the temperature the network becomes 
more connected; i.e. more particles become fully bonded. 
In Fig. \ref{fig2}  we present our results for the fractions of free and 3-times bonded particles,  $x_0$ and $x_3$, respectively, along the binodals at different strength of Yukawa interaction.
They are calculated using the following relation \cite{Wertheim1987}
\be
x_j=X^{n_s-j}(1-X)^j,
\label{fraction}
\ee
where $x_j$ is the fraction of $j$-times bonded particles.
Upon the temperature decrease, $x_3$ and $x_0$ monotonically approach their limiting values  $x_3\rightarrow 1$ and $x_0\rightarrow 0$ in the liquid phase and $x_3\rightarrow 0$ and $x_0\rightarrow 1$ in the gas phase, regardless of the number of patches and strength of Yukawa attraction. 
In Fig. \ref{fig3} we show our results for the RDFs $g_{01}(r)$ and $g_{11}(r)$ of the reference system. In the present case this is the hard-sphere (not the patchy hard-sphere) fluid confined in the Yukawa hard-sphere matrix. The RDFs are calculated at selected state points along the liquid phase binodal for the case with $\epsilon_{01}^{(Y)}/\epsilon_{11}^{(as)}=0.16$. Due to Yukawa attraction the contact value of the RDF $g_{01}(r)$ increases with the temperature decrease and as a consequence $g_{11}(\sigma_1)$ increases too. Corresponding changes in the shape of the RDF $g_{11}(r)$ reflect the fact that the fluid particles distribute themselves around the matrix obstacles close to their surfaces. Note again, that the RDFs shown here do not include the effects of the site-site square-well interaction between the fluid particles. An account for the presence of the square-well site-site bonding interaction between the fluid particles will not change 
this tendency it will merely contribute a high and narrow peak for the $g_{11}(r)$ at contact. 
On one side, the temperature decrease makes the network formed in the liquid phase to become more connected, while on the other it makes the fluid particles to distribute non-uniformly. Most of the fluid particles are located close to the obstacles and their global distribution is defined by the distribution of the matrix particles.
Finally, in Fig. \ref{fig4} we present our results for the difference in the entropy per particle  $\Delta S$ and internal energy per particle $\Delta U$ of the coexisting phases along the binodals for the 3- and 4-patch versions of the model at $\epsilon_{01}^{(Y)}/\epsilon_{11}^{(as)}=0.22$.
Similarly as in previous calculations \cite{Russo2011b} the curves for these differences follow
each other closely, with minor differences between models. They indicate that decrease in the energy of the liquid phase is compensated by the corresponding decrease in the entropy at given temperature. Such a behavior is  typical for the usual {colloids} 
but has been observed also for patchy colloidal fluids 
\cite{Russo2011b,Rovigatti2013}. 

{\it Conclusions.} --
We have shown that patchy colloidal model with three and four equivalent patches confined in the attractive random porous media undergo the re-entrant gas-liquid phase separation with the possibility for the liquid phase to have vanishing low density. This behavior is caused by an interplay between strong fluid-fluid bonding interaction and relatively weak fluid-matrix interactions. At high temperatures the shape of the phase diagram is 
defined by the patch-patch interaction between the fluid particles, while weak Yukawa attraction only slightly enhances the fluid-fluid bonding. At low temperature almost all the fluid particles are fully ($n_s$-times) bonded. Under such conditions the 
network of the fluid particles is formed  and the shape of the phase diagram is defined by the Yukawa fluid-obstacle attraction. Due to the fluid-matrix interaction, a layer of mutually bonded particles around the obstacles is formed and the corresponding 
network becomes strongly nonuniform. Distribution of the fluid 
particles in the network is defined by the  distribution of the obstacles. In such a situation all the particles in the gas phase are free, while those in the liquid are fully bonded. To maintain the mutual compensation of the energy and entropy changes upon phase separation, the liquid binodal moves in the direction of the lower densities, making the phase diagram re-entrant.

Our results provides further example of the peculiarities of the phase behavior of the patchy colloidal  fluids, which is currently a hot topic of the colloid science. In particular protein self-assembly and 
phase separation in crowded cell environment can be described using patchy colloid models confined in the porous media \cite{Hvozd2020}. Our study may shed some light on the very complicated phase behavior of biological macromolecules in crowded conditions. The nonuniform equilibrium gelation is of relevance for various biological studies \cite{Cai2017,Cai2018}. To quote \cite{Madl2017}: ``The proteins interact as patchy colloids and form an ordered fluid''. We hope that this and previous relevant studies will open new possibilities in 
making the equilibrium gels with the predefined nonuniform distribution of particles.\\

{\bf Acknowledgment:} 
YVK and TH gratefully acknowledge financial support from the National Research Foundation of Ukraine 
(project No.2020.02/0317). V.V. was supported by the Slovenian Research Agency fund (ARRS) through the 
Program 0103--0201 and the project J1--1708.

\begin{appendix}
\setcounter{equation}{0}
\renewcommand{\theequation}{A\arabic{equation}}

\section{Expressions for Helmholtz free energy, chemical potential and pressure of the
hard-sphere fluid confined in the hard-sphere matrix}

According to SPT~\cite{Holovko2009,Patsahan2011,Holovko2013}
\begin{equation}
{\beta\Delta A_{hs}\over N_{hs}}=\beta\Delta\mu_{hs}-{\beta \Delta P_{hs}\over\rho_{hs}}
\label{Ahs}
\end{equation}
and
\begin{equation}
g_{hs}={1\over \phi_0-\eta}+{3(\eta+\eta_0\tau)\over 2(\phi_0-\eta_0)^2},
\label{g}
\end{equation}
where $N_{hs}$ is the number of hard-sphere particles, i.e. $N_{hs}=7N$, $N$ is the number of the antibody molecules in the system, 
$$
{\beta\Delta P_{hs}\over\rho_{hs}}={1\over 1-\eta/\phi_0}{\phi_0\over\phi}
+\left({\phi_0\over\phi}-1\right){\phi_0\over\eta}\ln\left(1-{\eta\over\phi_0}\right)
$$
\begin{equation}
+{a\over 2}{\eta/\phi_0\over(1-\eta/\phi_0)^2}+
{2b\over 3}{(\eta/\phi_0)^2\over(1-\eta/\phi_0)^3}-1,
\label{P}
\end{equation}
$$
\beta\Delta\mu_{hs}=
\beta\mu^{(ex)}_{1}
-\ln\left(1-{\eta\over\phi_0}\right)
+
{\eta(\phi_0-\phi)\over\phi_0\phi(1-\eta/\phi_0)}
+\left(1+a\right){\eta/\phi_0\over(1-\eta/\phi_0)}
$$
\begin{equation}
+{(a+2b)\over 2}{(\eta/\phi_0)^2\over(1-\eta/\phi_0)^2}
+{2b\over 3}{(\eta/\phi_0)^3\over(1-\eta/\phi_0)^3},
\label{mu}
\end{equation}
and $\eta_0=\pi\rho_0\sigma_0^3/6$, $\phi_0=1-\eta_0$, $\eta=\pi\rho_{hs}\sigma_{hs}^3/6$
and $\phi=\exp{(-\beta\mu^{(ex)}_{1})}$.

\noindent
Here
\begin{equation}
a=6+{3\eta_0\tau\left(\tau+4\right)\over 1-\eta_0}+
{9\eta_0^2\tau^2\over(1-\eta_0)^2},
\;\;\;\;\;\;\;\;\;\;\;\;
b={9\over 2}\left(1+{\tau\eta_0\over 1-\eta_0}\right)^2,
\label{b1}
\end{equation}
$$
\beta\mu_{1}^{(ex)}=-\ln{(1-\eta_0)}
$$
$$
+{9\eta_0^2\over 2(1-\eta_0)^2}-\eta_0Z_0
+\left[3\eta_0Z_0-{3\eta_0(2+\eta_0)\over(1-\eta_0)^2}\right](1+\tau)
$$
\begin{equation}
-\left[3\eta_0Z_0-{3\eta_0(2+\eta_0)\over 2(1-\eta_0)^2}\right](1+\tau)^2
+\eta_0Z_0(1+\tau)^3,
\label{mu1}
\end{equation}
$Z_0=(1+\eta_0+\eta_0^2)/(1-\eta_0)^3$ and $\tau=\sigma_{hs}/\sigma_0$.\\

\section{Replica Ornstein-Zernike equations and their closures}

The set of the ROZ equations enables one to calculate the structure and thermodynamic properties of the fluid adsorbed into the disorder porous media \cite{Given1992,Given1993}. 
For  hard-sphere fluid confined in the Yukawa hard-sphere matrix the theory is represented 
by the OZ equation for the direct $c_{00}(r)$ and total $h_{00}(r)$ correlation functions,
describing the structure of the matrix, i.e.
\be
h_{00}-c_{00}=\rho_0c_{00}\otimes h_{00},
\label{MM}
\ee
and a set of three equations, which include direct $c_{01}(r),c_{11}(r),c_{11(1)}(r)$ and total
$h_{01}(r),h_{11}(r),h_{11(1)}(r)$ matrix-fluid (with the lower indices 01) and fluid-fluid 
(with the lower indices 11 and 11(1)) correlation functions, 

\be
h_{01}-c_{01}=\rho_0c_{01}\otimes h_{00}+\rho_1c_{11(1)}\otimes h_{10},
\label{01}
\ee
$$
h_{11}-c_{11}=\rho_0c_{10}\otimes h_{01}+\rho_1c_{11(1)}\otimes h_{11}
$$
\be
+\rho_1\left[c_{11}-c_{11(1)}\right]\otimes h_{11(1)},
\label{11}
\ee
\be
h_{11(1)}-c_{11(1)}=\rho_1c_{11(1)}\otimes h_{11(1)},
\label{111}
\ee
where the lower index $11(1)$ denote connectedness correlation functions,
 the symbol $\otimes$ denotes convolution.
This set of equations have to be supplemented by the corresponding closure relations. We are
using here hypernetted chain (HNC) approximation for the OZ equation (\ref{MM}) and Mean Spherical
Approximation (MSA) for the set of equations (\ref{01}), (\ref{11}) and (\ref{111}), i.e. 
\be
h_{00}(r)+1=e^{-\beta U_{hs}(r)+h_{00}(r)-c_{00}(r)}
\label{PY}
\ee
and
\be
c_{11}(r)=\left[t_{11}(r)+1\right]f_{11}^{(hs)}(r)
\label{MSA1}
\ee
\be
c_{01}(r)=-\beta U^{(Y)}_{01}(r)e_{01}^{(hs)}(r)+\left[t_{01}(r)+1\right]f_{01}^{(hs)}(r)
\label{MSA0}
\ee
\be
c_{11(1)}(r)=\left[t_{11}(r)+1\right]f_{11}^{(hs)}(r)
\label{MSA2}
\ee

Solution of the set of equations (\ref{MM}) - (\ref{111}) was obtained
numerically via direct iteration method.  Yukawa contribution to Helmholtz free energy 
$\Delta A_Y$ was calculated using the energy route, i.e.
\be
\beta\Delta A_Y= \int_0^\beta E_Yd\beta',
\label{AY1}
\ee
where for the excess internal energy $E_Y$ we have
\be
{\beta E_Y\over V}=2\pi\beta\rho_1\rho_0\int_0^\infty r^2
U^{(Y)}_{01}(r)g_{01}(r)dr,
\label{EY}
\ee

Corresponding contributions to the chemical potential and pressure are calculated using the 
standard thermodynamic relations.

\end{appendix}


\begin{thebibliography}{99}

\bibitem{Bianchi2006} E. Bianchi,J. Largo, P. Tartaglia, E. Zaccarelli, and F. Sciortino
Phys. Rev. Lett. {\bf 97}, 168301 (2006).
\bibitem{Tavares2009} J.M. Tavares, P.I.C. Teixeira, and M. M. Telo da Gama, Phys. Rev. E {\bf 80},
021506 (2009).
\bibitem{Russo2011a} J. Russo, J.M. Tavares, P.I.C. Teixeira, M.M. Telo da Gama, and F. Sciortino,
Phys. Rev. Lett. {\bf 106}, 085703 (2011).
\bibitem{Russo2011b} J. Russo, J.M. Tavares, P.I.C. Teixeira, M.M. Telo da Gama, and F. Sciortino,
J. Chem. Phys. {\bf 135}, 034501 (2011).
\bibitem{Heras2011} D. de las Heras, J.M. Tavares, and M.M. Telo da Gama,
J. Chem. Phys. {\bf 134}, 104904 (2011).
\bibitem{Kalyuzhnyi2013} Y.V. Kalyuzhnyi, and P.T. Cummings, J. Chem. Phys. {\bf 139}, 104905 (2013).
\bibitem{Rovigatti2013} L. Rovigatti, J.M. Tavares, and F. Sciortino, Phys. Rev. Lett. 
{\bf 111}, 168302 (2013).
\bibitem{Sokolowski2014} S. Sokolowski, and Y.V. Kalyuzhnyi, J. Phys. Chem. B {\bf 118}, 9076 (2014).
\bibitem{Tavares2017} J.M. Tavares, and P.I.C. Teixeira, Phys. Rev. E {\bf 95}, 012612 (2017).
\bibitem{Sciortino2017} F. Sciortino, and E. Zaccarelli, COCIS {\bf 30}, 90 (2017).
\bibitem{Ruzicka2011a} B. Ruzicka, E. Zaccarelli, L. Zulian, R. Angelini, M. Sztucki, A. Moussad,
T. Narayanan, and F. Sciortino, Nat. Mater. {\bf 10}, 56 (2011).
\bibitem{Ruzicka2011b} B. Ruzicka, and E. Zaccarelli, Soft Matter {\bf 7}, 1268 (2011).
\bibitem{Pujala2015} R.K. Pujala, N. Joshi, and H.B. Bohidar, Colloid Polym. Sci. {\bf 293}, 2883 
(2015).
\bibitem{Pajula2019} R.K. Pajula, and H.B. Bohidar, Colloid Polym. Sci. {\bf 297}, 1053 (2019).
\bibitem{Wertheim1987} M.S. Wertheim, J. Chem. Phys. {\bf 87}, 7323 (1987).
\bibitem{Holovko2009} M. Holovko, and W. Dong, J. Phys. Chem. B {\bf 113}, 6360 (2009).
\bibitem{Patsahan2011} T. Patsahan, M. Holovko, and W. Dong, J. Chem. Phys. {\bf 134}, 074503 (2011).
\bibitem{Holovko2013} M. Holovko, T. Patsahan, and W. Dong, Pure and Appl. Chem. {\bf 85}, 115(2013)
\bibitem{Given1992}  J.A. Given, and G. Stell, J. Chem. Phys. {\bf 97}, 4573 (1992).
\bibitem{Given1993}  J.A. Given, and G. Stell, Physica A {\bf 209}, 495 (1993).
\bibitem{Kalyuzhnyi2014} Y.V. Kalyuzhnyi, M. Holovko, T. Patsahan, and P.T. Cummings,
J. Phys. Chem. Lett. {\bf 5}, 4260 (2014).
\bibitem{Nelson2020} A.K. Nelson, Y.V. Kalyuzhnyi, T. Patsahan, C. McCabe, J. Molec. Liq. {\bf 300}, 
112348(2020).
\bibitem{Hvozd2020} T. Hvozd, Y.V. Kalyuzhnyi, V. Vlachy, Soft Matter {\bf 16}, 8432(2020).
\bibitem{Cai2017} J. Cai, J.P. Townsend, T.C. Dodson, P.A. Heiney, A.M. Sweeney, Science {\bf 357}, 564(2017).
\bibitem{Cai2018} J. Cai, A.M. Sweeney, ACS Cent. Sci. {\bf 4}, 840(2018).
\bibitem{Madl2017} T. Madl, Science {\bf 357}, 546 (2018).
\end{thebibliography}
\end{document}